\documentclass{article}

\usepackage{microtype}
\usepackage{graphicx}
\usepackage{subfigure}
\usepackage{multirow}
\usepackage{multicol}
\usepackage{booktabs}
\usepackage{xcolor}

\usepackage{hyperref}
\usepackage{url}
\usepackage{algorithm}
\usepackage{algpseudocode}
\usepackage{subcaption}
\usepackage{wrapfig}
\usepackage{amsthm}
\usepackage[T1]{fontenc}

\usepackage[accepted]{icml2023}

\usepackage{amsmath}
\usepackage{amssymb}
\usepackage{mathtools}
\usepackage{amsthm}

\usepackage[capitalize,noabbrev]{cleveref}

\theoremstyle{plain}

\theoremstyle{definition}

\theoremstyle{remark}

\usepackage[textsize=tiny]{todonotes}

\icmltitlerunning{Leveraging Team Correlation for Approximating Equilibrium in Two-Team Zero-Sum Games}

\begin{document}

\twocolumn[
\icmltitle{Leveraging Team Correlation for Approximating Equilibrium \\ in Two-Team Zero-Sum Games }

\begin{icmlauthorlist}
\icmlauthor{Naming Liu}{sjtu}
\icmlauthor{Mingzhi Wang}{pku}
\icmlauthor{Youzhi Zhang}{cair-cas}
\icmlauthor{Yaodong Yang}{pku}
\icmlauthor{Bo An}{ntu}
\icmlauthor{Ying Wen}{sjtu}

\end{icmlauthorlist}

\icmlaffiliation{sjtu}{Shanghai Jiao Tong University, Shanghai, China}
\icmlaffiliation{ntu}{Nanyang Technological University, Singapore}
\icmlaffiliation{pku}{Peking University, Beijing, China}
\icmlaffiliation{cair-cas}{Centre for Artificial Intelligence and Robotics, Hong Kong Institute of Science \& Innovation, Chinese Academy of Sciences, Hong Kong}
\icmlcorrespondingauthor{Youzhi Zhang}{youzhi.zhang@cair-cas.org.hk}
\icmlcorrespondingauthor{Ying Wen}{ying.wen@sjtu.edu.cn}

\icmlkeywords{Two-Team Zero-Sum Game, Unexploitable Equilibrium}

\vskip 0.3in
]

\printAffiliationsAndNotice{}

\begin{abstract}
Two-team zero-sum games are one of the most important paradigms in game theory. In this paper, we focus on finding an unexploitable equilibrium in large team games. An unexploitable equilibrium is a worst-case policy, where members in the opponent team cannot increase their team reward by taking any policy, e.g., cooperatively changing to other joint policies. As an optimal unexploitable equilibrium in two-team zero-sum games, correlated-team maxmin equilibrium remains unexploitable even in the worst case where players in the opponent team can achieve arbitrary cooperation through a joint team policy. However, finding such an equilibrium in large games is challenging due to the impracticality of evaluating the exponentially large number of joint policies. To solve this problem, we first introduce a general solution concept called restricted correlated-team maxmin equilibrium, which solves the problem of being impossible to evaluate all joint policy by a sample factor while avoiding an exploitation problem under the incomplete joint policy evaluation. We then develop an efficient sequential correlation mechanism, and based on which we propose an algorithm for approximating the unexploitable equilibrium in large games. We show that our approach achieves lower exploitability than the state-of-the-art baseline when encountering opponent teams with different exploitation ability in large team games including Google Research Football.

\end{abstract}

\section{Introduction}

The computational study of game theoretical solution concepts has been a central problem in Artificial Intelligence. For example, the study of Nash equilibrium \cite{NE} has achieved many progress in 2-player games including security games \cite{security} and poker games \cite{kuhn}. However, there are few results on two-team zero-sum (2t0s) games, even though the escalating prominence of real-world applications, such as Dota \cite{dota}, StarCraft \cite{starcraft1, starcraft2}, and Google Research Football \cite{football}, accentuates the imperative for advancing research on them. 

In the context of two-team zero-sum games, team members cooperatively compete against an adversary team, aiming to maximize their shared reward. This paper focuses on finding an unexploitable equilibrium in large scale two-team zero-sum games. An unexploitable equilibrium is a worst-case policy, where members in the opponent team cannot increase their team reward by deviation, e.g., cooperatively changing to other joint policies. Due to different correlations among teammates, the opponent team may have different exploitation abilities. For instance, when correlated by a joint team policy, members can cooperatively change to any joint policy after reaching an equilibrium, achieving optimal exploitation ability. In contrast, members without correlation cannot cooperatively change to other joint policies \cite{FCP}, resulting in the worst exploitation ability. Furthermore, members, who are correlated by specific cooperative methods (e.g., Multi-Agent PPO (MAPPO) \cite{mappo}), can only change their policies in accordance with the cooperative mechanisms.

With the correlation of joint team policy, Correlated-Team Maxmin Equilibrium (CTME) \cite{TME} becomes the optimal unexploitable equilibria in two-team zero-sum games. Once a CTME is reached, members in neither team have an incentive to cooperatively change their policies to any other joint policy. As a result, even in the worst-case where the opponent team has the optimal exploitability ability, CTME remain unexploitable. However, finding a CTME is impractical in large team games. This is because, in large games, it is impossible to evaluate all the exponential large number of team joint policies. Existing algorithms for large two-team zero-sum games also cannot find an unexploitable equilibrium in large team games. To extend to large two-team zero-sum games, 
Team-Policy Space Response Oracles (Team-PSRO) \cite{TeamPSRO} is proposed to approximate TMECor \cite{CG}, a CTME in imperfect information extensive form games. However, under the synchronized mechanism of MAPPO, the team policy search space is restricted to a synchronized subset of joint policy space, making the equilibrium found by Team-PSRO exploitable by the opponent team who can cooperatively change to a non-synchronized joint policy. As a result, Team-PSRO fails to converge to an unexploitable equilibrium in large two-team zero-sum games.

While there exist other equilibria \cite{TME1997, TME, TMEconverging} in two-team zero-sum games, these equilibria are either hard to find or prone to be exploited by a cooperating team. Nash equilibrium, which is an unexploitable policy in numerous scenarios including 2-player games, may get exploited by a cooperating team in two-team zero-sum games. This is because, although no player can increase the team reward by unilaterally changing her individual policy, the players within the same team, who can get correlated by information sharing, may be motivated to cooperatively change their policies for a higher team reward \cite{FCP}. Although team maxmin equilibrium \cite{TME1997, TMEconverging} assumes the opponent team is correlated by a joint adversary policy, thus overcoming the exploitation problem in NE, it is hard to scale to large games due to the high computational complexity \cite{TME, CG}. As a result, how to find an unexploitable equilibrium in large team games remains an open challenge. 

To tackle these challenges, we first introduce a general solution concept, called restricted Correlated-team maxmin equilibrium (rCTME), which encompasses existing equilibria in team games by flexibly scaling the deviation policy space. By restricting the growth rate of joint policies to be evaluated with a sample factor, rCTME solves the problem of being impossible to evaluate all joint policies in large games, while avoiding an exploitation problem under the incomplete joint policy evaluation. We then develop a new sequential correlation mechanism and propose an efficient algorithm S-PSRO to find the unexploitable rCTME with sequential correlation in large games. By scaling the sample factor in different game scales, S-PSRO strikes a balance between exploitation and efficiency. We show that our technique can achieve lower exploitability than the state-of-the-art baseline when confronting opponent teams with different exploitation ability in large team games including Google Research Football \cite{football}.

\section{Preliminaries}
\label{sec: formulation}

\subsection{Two-Team Zero-Sum Game}
Consider a two-team stochastic game \cite{markov}\footnote{Our methods mostly apply to stochastic games including Google Research Football mentioned in Section~\ref{sec:grf}. Normal-form games can be considered as a special case of stochastic games with $|\boldsymbol{\mathcal{O}}| = 1$.} G = $(\mathcal{T}, \boldsymbol{\mathcal{O}}, \boldsymbol{\mathcal{A}}, R, P, \gamma)$.
We denote the set of two teams as $\mathcal{T} = \{T_1, T_2\}$, where $T_j = \{1,2, ..., n_j\}$ is the finite set of teammates; 
 and $\boldsymbol{\mathcal{O}} = \times_{j=1}^2\boldsymbol{\mathcal{O}_j}$ is the product of local observation space of two teams, namely the joint observation space, where $\boldsymbol{\mathcal{O}_j} = \times_{i=1}^{n_j} O_j^i$ is the product of local observation space of the teammates in team $j$, namely team $j$'s joint observation space. Let $\boldsymbol{\mathcal{A}} = \times_{j=1}^2 \boldsymbol{\mathcal{A}_j}$ be the product of action space of two teams, namely the joint action space, where $\boldsymbol{\mathcal{A}_j} = \times_{i=1}^{n_j} A_j^i$ is the product of action space of teammates in team $j$, namely team $j$'s joint action space. In addition, $R_j: \boldsymbol{\mathcal{O}} \times \boldsymbol{\mathcal{A}} \rightarrow [-R_{max}, R_{max}]$ is the joint reward function of team $j$. Let $P: \boldsymbol{\mathcal{O}} \times \boldsymbol{\mathcal{A}} \times \boldsymbol{\mathcal{O}} \rightarrow \mathbb{R}$ is the transition probability function, and $\gamma \in [0, 1)$. At time step $t\in \mathbb{R}$, a team $j \in \{1, 2\}$ observes an observation $o_{j, t} \in \boldsymbol{\mathcal{O}_j}$ ($o_{j, t} = (o_{j, t}^1, ..., o_{j, t}^{n_j})$ is a "joint" observation) and takes a team joint action $a_{j, t} \in \boldsymbol{\mathcal{A}_j}$ according to its policy $\boldsymbol{\pi_j}$, where $\boldsymbol{\pi_j} = (\pi_{j,1}, ..., \pi_{j,n_j})$. At each time step, two teams take actions simultaneously based on their observations with no sequential dependency. At the end of each time step, the team $j$ receives a joint reward $R_j(o_{0, t}, a_{0, t}, o_{1, t},  a_{1, t})$ and observes $o_{j, t+1}$. Following this process infinitely long, the team $j$ earns a discounted cumulative return of $R^{\gamma}_j \triangleq \Sigma_{t=0}^{\infty} \gamma^tR_j(o_{0, t}, a_{0, t}, o_{1, t},  a_{1, t})$.

In this paper, we consider two-team zero-sum games, where players within the same team are fully cooperative and
share the same utility. Let $R_{j, n}, j \in \{1, 2\}, n \in T_j$ denote the reward function of player $n$ in team $j$, we have 
$$R_{j, 1} = ... = R_{j, n_j} = R_j.$$
On the other hand, the two teams are fully competitive and their rewards sum to zero, i.e.,
$$R_1 + R_2 = 0.$$

\subsection{Nash Equilibrium}
Nash equilibrium \cite{NE} is a stable policy in many scenarios including 2-player games. Upon reaching a Nash equilibrium, no player has an incentive to change her policy given other players’ policies (e.g., other teammates and all opponents). However, as shown in Example 1, in two-team zero-sum games Nash equilibrium policies may get exploited by a cooperating opponent team.

\textbf{Example 1.} Consider a normal form game where each team has two players $T_1 = T_2 = \{1,2\}$, and each player has two actions, $A_1^1 = A_1^2 = A_2^1 = A_2^2 = \{0,1\}$. $\pi_{j,n}^{(a)}$ represents the probability of action $a$ for the player $n$ in team $j$. The reward functions are defined as follows:
\begin{align*}
 R_1 &= \left\{ 
    \begin{aligned}
    &2 & \pi_{1, 1}^{(1)}= \pi_{1, 2}^{(1)}= 1, \pi_{2, 1}^{(0)} = \pi_{2, 2}^{(0)} = 1 \\
    &1 + \nu_2 - \nu_1 & otherwise
    \end{aligned}
  \right. \\
 R_2 &= - R_1,~ \text{where}: \\
 \nu_1 &= \pi_{1, 1}^{(1)} * 2 + \pi_{1, 2}^{(1)}, 
 \nu_2 = \pi_{2, 1}^{(1)} * 2 + \pi_{2, 2}^{(1)}.
\end{align*}
The above game has a Nash equilibrium policy $((1, 0), (1, 0), (1, 0), (1, 0))$, where all players choose action 0 with probability 1. Rewards of this policy are $R_1 = 1, R_2 = -1$. However, this Nash equilibrium policy might be exploited by a cooperating team. Concretely, if the players in $T_1$ are able to cooperatively change their policies, they can be motivated to deviate from $(0, 0)$ to $(1, 1)$ and obtain a higher reward $R_1 = 2$.

In this paper, we focus on finding an unexploitable equilibrium, which is a worst-case policy, where members in the opponent team cannot increase their team reward by taking another policy, e.g., cooperatively changing to other joint policies, in large two-team zero-sum games.

When confronting an opponent team, an equilibrium is exploited when the opponent players cooperatively change to other joint policies and get a higher reward, as a result lowering the reward of the other team. Different correlations among members lead to different exploitation ability of the opponent team. For instance, when correlated by a joint team policy, members can cooperatively change to any joint policy after reaching an equilibrium, achieving optimal exploitation ability. In contrast, members without correlation cannot cooperatively change to any joint policy \cite{FCP}. Whether an equilibrium is exploitable is closely related to the exploitation ability of the opponent team.

On the other hand, an unexploitable equilibrium is a policy, where changing the opponents' policy to the joint policies that they can change to cannot make their reward higher. To find an equilibrium that is unexploitable when confronting any opponent team, the optimal method is to assume that the opponent team has optimal exploitation ability.

\subsection{Correlated-Team Maxmin Equilibrium (CTME)}
\label{sec: teal-level NE}
By correlating members with a joint team policy, Correlated-team maxmin equilibrium \cite{TME} becomes an optimal unexploitable equilibrium in two-team zero-sum games. Upon reaching such an equilibrium, members in neither team can increase their team reward by cooperatively change their policies to any other joint policy. Formally, a policy $(\boldsymbol{\pi_1}^*, \boldsymbol{\pi_2}^*)$ is a CTME if, for each team $T_j, j\in \{1, 2\}$, the team joint policy $\boldsymbol{\pi_j}^*$ is a best response to the opponent team's policy $\boldsymbol{\pi_{-j}}^*$, formally,
\begin{equation}
\label{eq:ctme1}
R_j(\boldsymbol{\pi_{j}}^*, \boldsymbol{\pi_{-j}}^*) \geq R_j(\boldsymbol{\pi_{j}}, \boldsymbol{\pi_{-j}}^*) \quad \forall \boldsymbol{\pi_j} \in \boldsymbol{\Pi_j},
\end{equation}
where $\boldsymbol{\Pi_j}$ is the joint policy set of $T_j$.

As a result, even in the worst case where
the opponent team has the optimal exploitability ability and is able to change to any other joint policy, CTME remain unexploitable.

However, finding a CTME in large-scale team games is impractical. This is because, as shown in equation~\ref{eq:ctme1}, obtaining a CTME requires to compare the reward of the equilibrium candidate with all other joint policies in both teams. When it comes to large scale team games where both teams' joint policy space is exponentially large, it is impossible to evaluate all the exponentially large number of joint policies. For example, when two teams both have $\boldsymbol{t}$ players, each with $\boldsymbol{n_a}$ actions and $\boldsymbol{n_o}$ possible private observations, then the number of pure joint policies of both teams is $(n_a^{\boldsymbol{t}})^{({\boldsymbol{n_o^t}})}$, which grows exponentially with increase of $\boldsymbol{t}$ and $\boldsymbol{n_o}$. In this context, a CTME is a distribution, over the exponentially large space  $(n_a^{\boldsymbol{t}})^{(\boldsymbol{n_o^t})\times 2}$, that satisfies equation~(\ref{eq:ctme1}), which requires to compare with $2\times(n_a^{\boldsymbol{t}})^{(\boldsymbol{n_o^t})}$ policies. Searching a distribution over an exponential large space, which has higher reward than $2 \times (n_a^{\boldsymbol{t}})^{({\boldsymbol{n_o^t}})}$ policies, is impractical.

As a result, how to find an unexploitable equilibrium in large-scale two-team zero-sum games remains an open challenge.

\section{Restricting Equilibrium via Team Correlation}
An unexploitable equilibrium in two-team zero-sum games requires the team correlation assumed by the equilibrium has higher exploitation ability than the opponent team's exact correlation.

As shown above, finding the optimal unexploitable equilibrium CTME in large games is impractical. This is because, finding a CTME, where members are correlated by joint policies, requires to compare the reward of each equilibrium candidate with two teams' joint policies. When it comes to large games, it is impossible to evaluate all the exponentially large number of joint policies. At the same time, it is also hard for members in the opponent team to achieve the optimal exploitation ability due to the same reason.

To address the above challenges and find an unexploitable equilibrium in large scale two-team zero-sum games, we propose a general solution concept named restricted CTME, which encompasses existing equilibria in team games by flexibly  scaling the deviation policy space. rCTME solves the problem of being impossible to evaluate all joint policies by restricting joint policies to be evaluated through a sample factor while avoiding exploitation problem under the incomplete joint policy evaluation.

\subsection{Deviation Policy Space}
Before providing the formal definition of rCTME, we first introduce the concept of deviation policy space. 

\textbf{Definition 1 (Deviation Policy Space)} A deviation policy space is the set of policies that players or teams can transition to after reaching an equilibrium. If the rewards of policies in the deviation policy space are all less than or equal to the reward of $\pi^*$, we conclude that $\pi^*$ is an equilibrium corresponding to this deviation policy space. Different equilibria have different deviation policy space. For example, the deviation policy space for team $T_j$ in NE is 
$$\cup_{n\in T_j} \{(\pi_{j, n}, \pi_{j, -n}^*)| \forall \pi_{j, n} \in \Pi_{j, n}\}, $$ 
where $\pi_{j, n}$ represents the policy of player $n$, $\pi_{j, -n}^*$ represents the equilibrium policy candidate to be evaluated for players other than player $n$, $\Pi_{j, n}$ is player $n$'s policy space, and the deviation policy space of team $T_j$ in CTME is, 
$$\{\boldsymbol{\pi_j}| \forall \boldsymbol{\pi_j} \in \boldsymbol{\Pi_j}\},$$
where $\boldsymbol{\pi_j}$ represents the team joint policy, $\boldsymbol{\Pi_j}$ is the joint policy space of team $T_j$.

We further divide the deviation policy space into two parts:
The deviation policy space of player $n$ in team $T_j$ is a tuple $ (I_{j, n}, C_{j, n})$, where $I_{j, n} \subseteq \{(\pi_{j, n}, \pi_{j, -n}^*)| \forall \pi_{j, n} \in \Pi_{j, n}\}$ is the individual deviation policy space, and $C_{j, n}\subseteq{\boldsymbol{\Pi_j}}$ is the correlated deviation policy space, which is enabled by the correlation among teammates. Note that $C_{j, n}$ is a subset of $\boldsymbol{\Pi_{j}}$ and the size of $C_{j, n}$ is associated to the specific correlation mechanism. We use the pure policy subset of correlated deviation policy space $|\cup_{i\in T_j}P(C_{j, n})|$ to measure the cooperative ability.

\subsection{Restricted CTME}
Based on the deviation policy space defined above, we propose a new solution concept, restricted Correlated-team maxmin equilibrium.

\textbf{Definition 2 (Restricted Correlated-Team Maxmin Equilibrium)}
Under a specific correlation mechanism, a restricted Correlated-team maxmin equilibrium is reached if no player can increase the team reward by changing her policy to policies in its deviation policy space $(I_{j, n}, C_{j, n})$. Formally, for each player $n \in T_j, \forall j\in\{1, 2\}$, 

\begin{equation}
\label{eq:coe}
\begin{aligned}
R_j(\pi_{j, n}^*, \pi_{j, -n}^*, \boldsymbol{\pi_{-j}}^*) \geq R_j(\pi_{j, n}, &\pi_{j, -n}^*, \boldsymbol{\pi_{-j}}^*)\\  &\forall \pi_{j, n}\in I_{j, n},
\end{aligned}
\end{equation}
\begin{equation}
\label{eq:coe1}
\begin{aligned}
R_j(\pi_{j, n}^*, \pi_{j, -n}^*, \boldsymbol{\pi_{-j}}^*) \geq  R_j&(\pi_{j, n}, \pi_{j, -n}, \boldsymbol{\pi_{-j}}^*) \\
 &\forall (\pi_{j, n}, \pi_{j, -n})\in C_{j, n}.
\end{aligned}
\end{equation}
Upon reaching an rCTME, the opponent team cannot get higher reward by cooperatively changing to policies within the corresponding deviation policy space. As a result, rCTME avoids being exploited under the incomplete joint policy evaluation.

By flexibly scaling deviation policy space, rCTME encompasses existing equilibria in 2t0s games. Table~\ref{tab:deviation set} displays the deviation policy space for these equilibria.

\subsection{Sample Factor}
As the game scales up, it becomes impractical to compare the reward of a candidate equilibrium policy with all joint policies. rCTME addresses this problem by applying a dynamic sample factor to the deviation policy space to limit the exponential growth of evaluated deviation policies. 

\begin{table*}[t]
\caption{Equilibrium Comparisons of NE, CTME and rCTMEs under different correlation mechanisms. rCTME$^1$ and rCTME$^2$ represent rCTME under sequential correlation with flexible sample factor and rCTME under pivot-followers correlation, a specific sequential correlation respectively. Both correlation mechanisms can be found in Section~\ref{sec:consensus}. Note that NE is a special rCTME under No Correlation Consensus. $P(\boldsymbol{\Pi_j})$ represents a pure policy subset of $\boldsymbol{\Pi_j}$. $I_j$ represents the individual deviation policy space. The deviation policy space of rCTME$^2$ is related to the sample factor. Equilibria with higher cooperative ability will be more resilient to the exploitations of the opponent team. In large scale games, rCTME$^2$ can be found more efficiently than CTME because with restrictions of the sample factor, its deviation policy space (or equivalent constraints) grows linearly, while that in CTME grows exponentially.}
\label{tab:deviation set}
\vskip 0.15in
\begin{center}
\begin{small}
\begin{sc}
\scalebox{0.8}{
    \begin{tabular}{ccccc}
        \toprule
        Equilibrium & Correlation Method &Equilibrium Search Space& Deviation Policy Space & Cooperative Ability  \\
        \midrule
        CTME & Centralized Correlation Device & $\times_{j\in{1, 2}}\boldsymbol{\Pi_j}$ & $\{\boldsymbol{\pi_j}|\forall \boldsymbol{\pi_j} \in \boldsymbol{\Pi_j} \}$ & $|P(\boldsymbol{\Pi_j})| - |I_j|$  \\
        
        NE &No Correlation& $\times_{j\in \{1,2\}, n\in T_j}\Pi_{j,n}$ & $\cup_{n\in T_j} \{(\pi_{j, n}, \pi_{j, -n}^*)| \forall \pi_{j, n} \in \Pi_{j, n}\}$& 0 \\
        rCTME$^1$& Pivot-follower Correlation & $\times_{j\in \{1,2\}} \Pi_{j, p_j}$ &$\{(\pi_p, ..., \pi_p) | \forall \pi_p \in \Pi_{j, p_j}\}$ & $|P(\Pi_{j, p})|$ \\
        
        \multirow{2}*{rCTME$^2$} & \multirow{2}*{Sequential Correlation} & \multirow{2}*{$\times_{j\in \{1, 2\}}\boldsymbol{\Pi_j}$} & \multirow{2}*{A flexible subset of $\boldsymbol{\Pi_j}$}& $N_{init} + \delta T \times f_T - |I|$ \\
        &&&&$N_{init} + \delta \pi \times f_{\pi} - |I|$ \\
        \bottomrule
    \end{tabular}
    }
\end{sc}
\end{small}
\end{center}
\vskip -0.1in
\end{table*}

\textbf{Definition 3 (Sample Factor)} Because the evaluated deviation policies grows at different rates when increasing the number of teammates in $T$ and the individual policy $\mathcal{\pi}$, the sample factor is defined a pair ($f_T$, $f_\pi$), where $f_T \in [0, C\times N_{init}]$ represents the growth rate of evaluated deviation policies when increasing $\delta T$ teammates in $T$, and $f_\pi \in [0, N_{init}]$ represents the growth rate of evaluated deviation policies when increasing $\delta \pi$ policies in $\pi$. Suppose there are $N_{init}$ deviation policies to be evaluated before increasing $T$ or $\pi$, and after an increase in $T$ or $\pi$, there are $N$ deviation policies to be evaluated,
$$N = N_{init} + \delta T \times f_T,$$ 
$$N = N_{init} + \delta \pi \times f_{\pi}.$$

With the restriction of sample factor, the deviation policies to be evaluated no long experience the exponential growth in large games. For example, with sample factor $f_T=100$, when increasing the teammates in $T$ from 10 to 100, where each player has 10 actions, deviation policies to be evaluated grows from $10^{10}$ to $10^{10} + 9\times10^3$, while without sample factor, the number of deviation policies becomes $10^{20}$. Then if we continue to increase the number of actions of players from 10 to 1000, with sample factor $f_\pi = 100$, the number of deviation policies grows from $10^{10}+9\times10^3$ to $10^{10}+9^3+9.9\times10^4$. While without sample factor, the number of deviation policies becomes $10^{2000}$.

\begin{figure}[ht!]
    \centering
    \includegraphics[width=0.75\columnwidth]{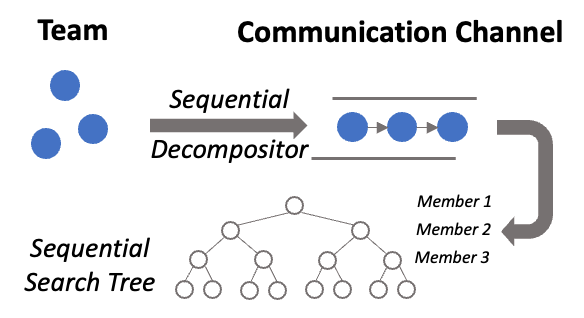}
    \caption{Mechanism of Sequential Correlation, under which the team policy search process is organized as a sequential search tree. The sequential information sharing is implemented through a communication channel.}
    \label{fig: decompositor}
\end{figure}

\section{Sequentially Correlated Equilibrium}
Under the framework of rCTME, we propose an efficient correlation mechanism with the flexible sample factor. We then compare the sequential rCTMEs with specific sample factors as well as their complexity. Last, based on sequential correlation, we propose an algorithm to approximate rCTME in large scale two-team zero-sum games.

\label{sec:consensus}

\subsection{Sequential Search Tree}

The mechanism of sequential correlation is decomposed into a sequential search tree. As shown in figure~\ref{fig: decompositor}, to construct a sequential search tree, a \emph{sequential decompositor} first decomposes players within the same team, who make synchronous decisions, into an agent-by-agent communication channel. Then the policy search process for these players is organized as a sequential search tree, where each non-terminal node represents an updated individual policy outcome of the corresponding player, each terminal node represents an updated team policy outcome consisting of correlated individual policies, each edge represents a transmission process of individual policies from the preceding node. 

For the example depicted in figure~\ref{fig: decompositor}, during the policy search phase, Mem 1 initiates the policy update process and transmits the updated policy to Mem 2. Based on Mem 1's policy, Mem 2 updates its policy and transmits both its own and Mem 1's policy to Mem 3. Subsequently, Mem 3 updates its policy based on the policies of Mem 1 and Mem 2. Eventually, the different combinations of players' policy updates lead to different terminal nodes. In this way, the sequential search tree enables subsequent players to update their policies based on the decision preferences of preceding players, thus fostering internal collaboration within the team. During the decision-making phase, players make decisions based on policies that incorporate the preferences of their teammates, and achieve spontaneous collaboration without additional in-game communication. 

While the sequential correlation does not directly restrict the exponential search space, it effectively addresses the issue of random search in Nash equilibrium through sequential correlation, thereby enhancing search efficiency. Additionally, this correlation mechanism imposes a flexible restriction on the growth of constraints through the sample factor.

\begin{figure*}
    \centering
    \includegraphics[scale=0.31]{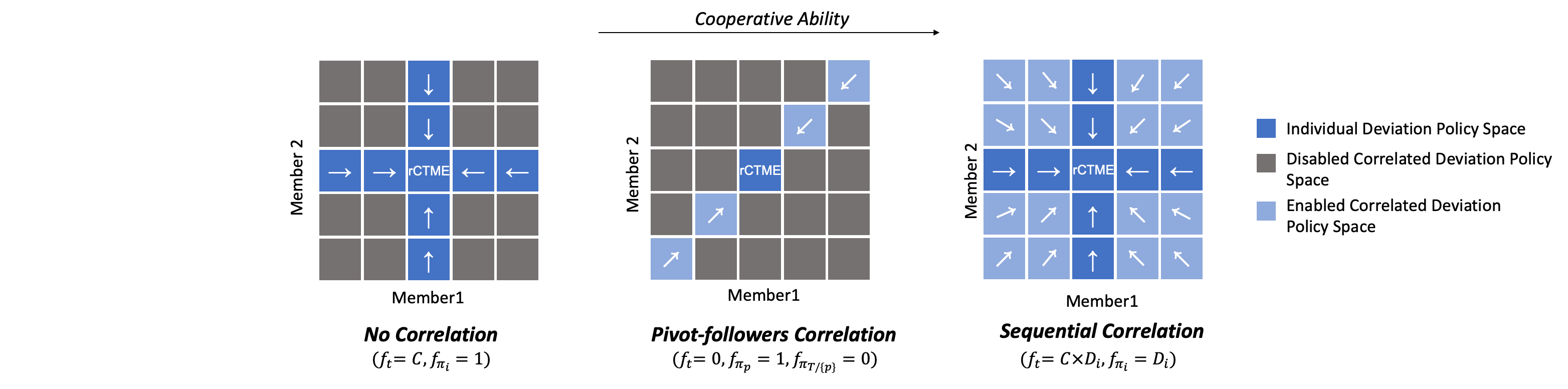}
    \caption{Deviation Policy Space Comparison of rCTME under Different Sequential Correlation Mechanisms, where No Correlation and Pivot-followers Correlation are two examples of Sequential Correlation. In the scenario depicted above, both teams have two players, referred as member 1 and member 2. The horizontal and vertical axes represent actions (policies) of the two players, the squares in the two-dimensional space indicating the team joint actions (policies), and the direction of the arrows indicates an increase in the joint action (policy) reward. Due to the symmetrical setup of the two teams, we only illustrate the deviation policy space of one team.}
    \label{fig: comparison}
\end{figure*}

\subsection{Sequential rCTMEs with Specific Sample Factor}
In this section, we introduce two sequential correlation mechanisms with specific sample factor: No Correlation and \textit{Pivot-followers} Correlation, and compare the rCTMEs under these mechanisms. The comparison of rCTME under these correlation mechanisms is shown in figure~\ref{fig: comparison}.

\textbf{No Correlation} As a special rCTME, the correlation in NE assumes that players within the same team cannot communicate or collaborate. Consequently, players neglect the received policies and update their policies independently. The lack of correlation leads to an exponentially growing search space and linearly growing constraints. The exponential growing search space is attributed to the fact that each player can freely choose policies within their policy space, resulting in the possibility of attaining any team policy. The linearly growing constraints is attributed to the empty correlated deviation policy space. As shown in figure~\ref{fig: comparison}, the number of constraints in No Correlation grows linearly both when increasing teammates and their action space. 

From the example in figure~\ref{fig: comparison}, it is evident that for each player in the team, the individual deviation policy space is equal to the individual policy space due to their independence, formally,
$$I_{1, 1} = \{(\pi_{1, 1}, \pi_{1, 2}^*)| \forall \pi_{1, 1} \in \Pi_{1, 1}\},$$
$$I_{1, 2} = \{(\pi_{1, 1}^*, \pi_{1, 2})| \forall \pi_{1, 2} \in \Pi_{1, 2}\},$$

and the correlated deviation policy space is empty due to the lack of correlation, formally,
$$C_{1, 1} = \emptyset, C_{1, 2} = \emptyset.$$

The cardinality of deviation policy space in NE is $|I_{1, 1}| + ... + |I_{1, n_1}|$, which is further equal to $|\Pi_{1, 1}| + 
... + |\Pi_{1, n_1}|$. As a result, with an increase of the number of teammates and policy space, the number of constraints only grows linearly. In this way, though No Correlation mechanisms imposes no restriction on the search space, it restricts the explosion of the constraints.

\textbf{Pivot-Followers} Under the \emph{pivot-followers} correlation, all members within the team make decisions based on the pivot \textit{p}'s individual policy $\pi_p$, resulting a synchronized correlation. The synchronized-level correlation ensures that both the number of the search space and constraints remain constant with an increase in teammates and grow linearly with an increase of the pivot's policy space. The growth of search space is attributed to the \emph{pivot-followers} consensus, where a pivot is elected to represent the entire team, transforming the two-team zero-sum game into a two-pivot zero-sum game (two-player zero-sum game). As a result, the search space is restricted to the pivot's individual policy space, whose size remains constant, regardless of the number of teammates in the team. Upon reaching the rCTME under the \emph{pivot-followers} consensus, any change in player $i$'s policy must incur identical changes in the policies of all $i$'s teammates. Consequently, the individual deviation policy space for each player in this case is empty, and the correlated deviation policy space for each player consists of homogeneous policies of teammates. The number of policies in the correlated deviation policy space is equal to the size of the pivot's individual policy space, making the growth of constraints being the same as the growth of the search space.

\subsection{Complexity of Sequential rCTME}
rCTME can be obtained by solving the linear programs in equations~(\ref{eq:coe}) and (\ref{eq:coe1}). The complexity of finding an rCTME under sequential correlation is closely related to the selection of sample factor. Without the constraint of sample factor, the complexity of finding an rCTME under sequential correlation is equal to the complexity of finding an CTME. The complexity of rCTME under specific sequential correlations is as follows.

\textbf{No Correlation} rCTME under no correlation consensus is an NE. In this setting, players perceive both opponents and teammates indiscriminately, making the 2t0s game indistinguishable from a multiplayer game. As a result, finding an NE in two-team zero-sum games is equal to find an NE in multiplayer games. Approaches \cite{multiplayerNE} in multiplayer games can thus be incorporated. It is important to note that the time complexity of finding an NE in the adversarial team game \cite{CG, TME}, a special 2t0s games, is CLS-hard \cite{complexity}.

\textbf{Pivot-Followers} \emph{pivot-followers} correlation elects a pivot to represent the entire team, transforming a two-team zero-sum game into a two-pivot zero-sum game (two-player zero-sum game). As a result, the complexity of finding an rCTME under \emph{pivot-followers} consensus in two-team zero-sum games is equal to the complexity of finding an NE in the corresponding two-pivot zero-sum games (two-player zero-sum game).

In large scale two-team zero-sum games, McAleer et al. ~(\citeyear{TeamPSRO}) introduce Team-PSRO to approximate TMECor \cite{CG}, a CTME in the imperfect information extensive form games. Team-PSRO correlates teammates by a parameter sharing based technique MAPPO \cite{mappo},  achieving a synchronized-level correlation. Due to the restrictions imposed on team joint policy space and constraints, Team-PSRO fails to converge to TMECor. Rather, Team-PSRO converges to the rCTME under pivot-followers consensus.

\subsection{Practical Algorithm: Sequential PSRO}
In this section, we introduce an approximate algorithm named Sequential Policy-Space Response Oracle (S-PSRO) to approximate the sequential rCTME in large scale games. 

S-PSRO implements the sequential correlation mechanism through the multi-agent advantage decomposition lemma \cite{happo}. Based on this lemma, teammates sequentially update their individual policies and are guaranteed to achieve a monotonic improvement on the team reward. As the max training iteration increases, the sample factor in the sequential correlation consensus also increases. This configuration enables adjusting the sample factor to strike a balance between training efficiency and equilibrium payoffs across different game scales.
\label{sec: s-psro}

S-PSRO iteratively trains a population of team policies composed of sequentially correlated individual policies. S-PSRO starts with a randomly initialized population Pop. At each iteration, it employs SeBR to determine the best response to the meta-policy of population Pop and then adds the derived best response into Pop. SeBR is a concrete implementation of sequential correlation cooperative consensus. More details on SeBR can be found in the Appendix.~\ref{sec: s-psro}.

The pesudocode of S-PSRO is shown in Algorithm~\ref{algo:spsro}. 

\begin{algorithm}
\caption{S-PSRO}
\label{algo:spsro}
\begin{algorithmic}[1]
\State Initialize the population as $\mathcal{P}$
\State $\mathcal{C} \gets \text{CommChannel()}$ \Comment{Communication Channel}
\For{$i = 1, \dots, \text{MaxIter}$}
    \State $\text{MetaPolicy} \gets \text{MetaSolver}(\mathcal{P})$ \Comment{Meta Solver}
    \State $br \gets \text{SeBR}(\text{MetaPolicy}, Q)$ \Comment{Sequential Best Response Oracle (SeBR)}
    \State $\mathcal{P}.\text{append}(br)$
\EndFor
\State \textbf{return} $\text{MetaSolver}(\mathcal{P})$
\end{algorithmic}
\end{algorithm}

\begin{figure*}[htbp]
    \centering
    \subfigure[Elo Ratings]{
        \includegraphics[width=0.42\columnwidth]{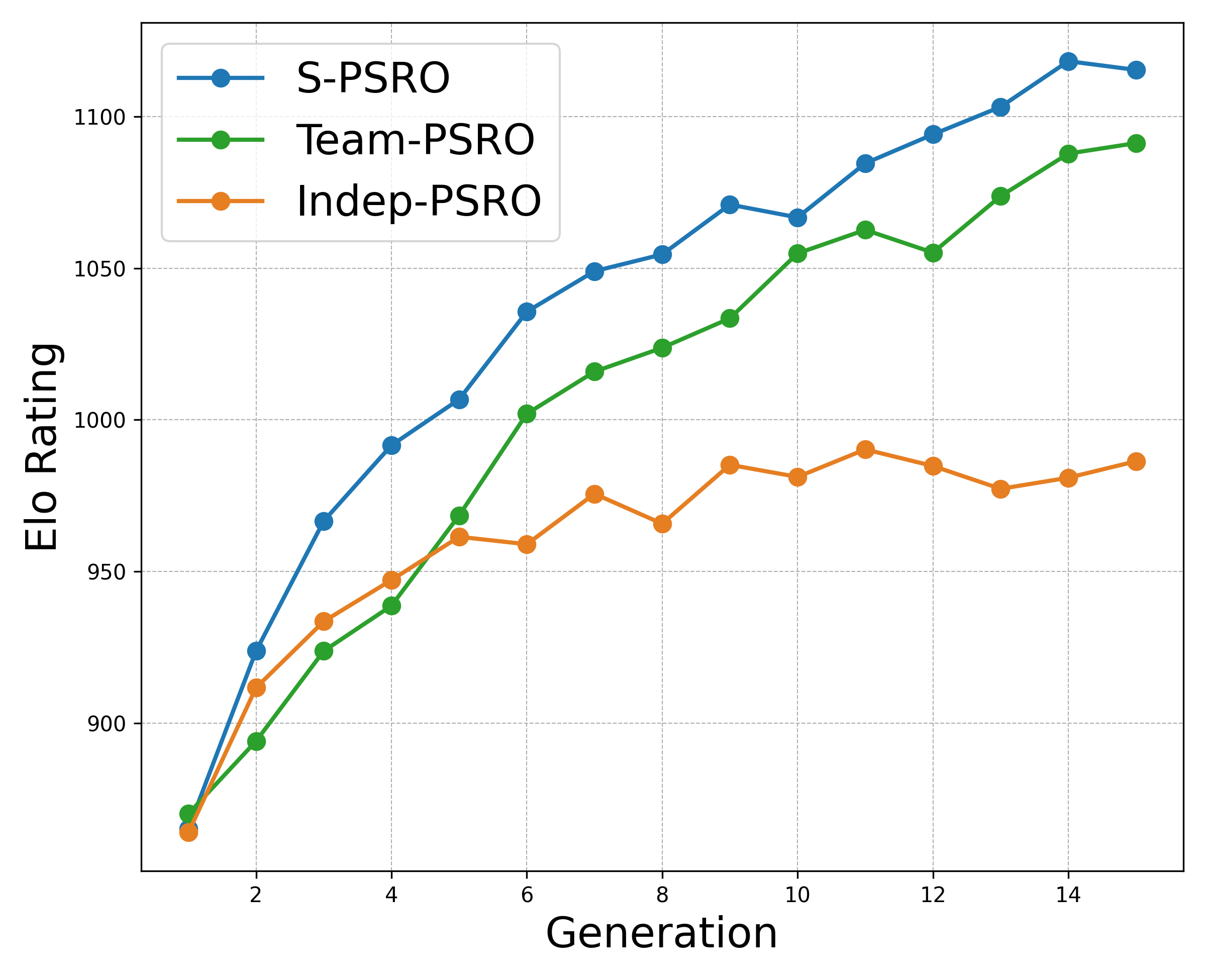}
        \label{fig:gd}
    }
    \subfigure[Goal Diff. vs. Built-in AI] {
        \includegraphics[width=0.46\columnwidth]{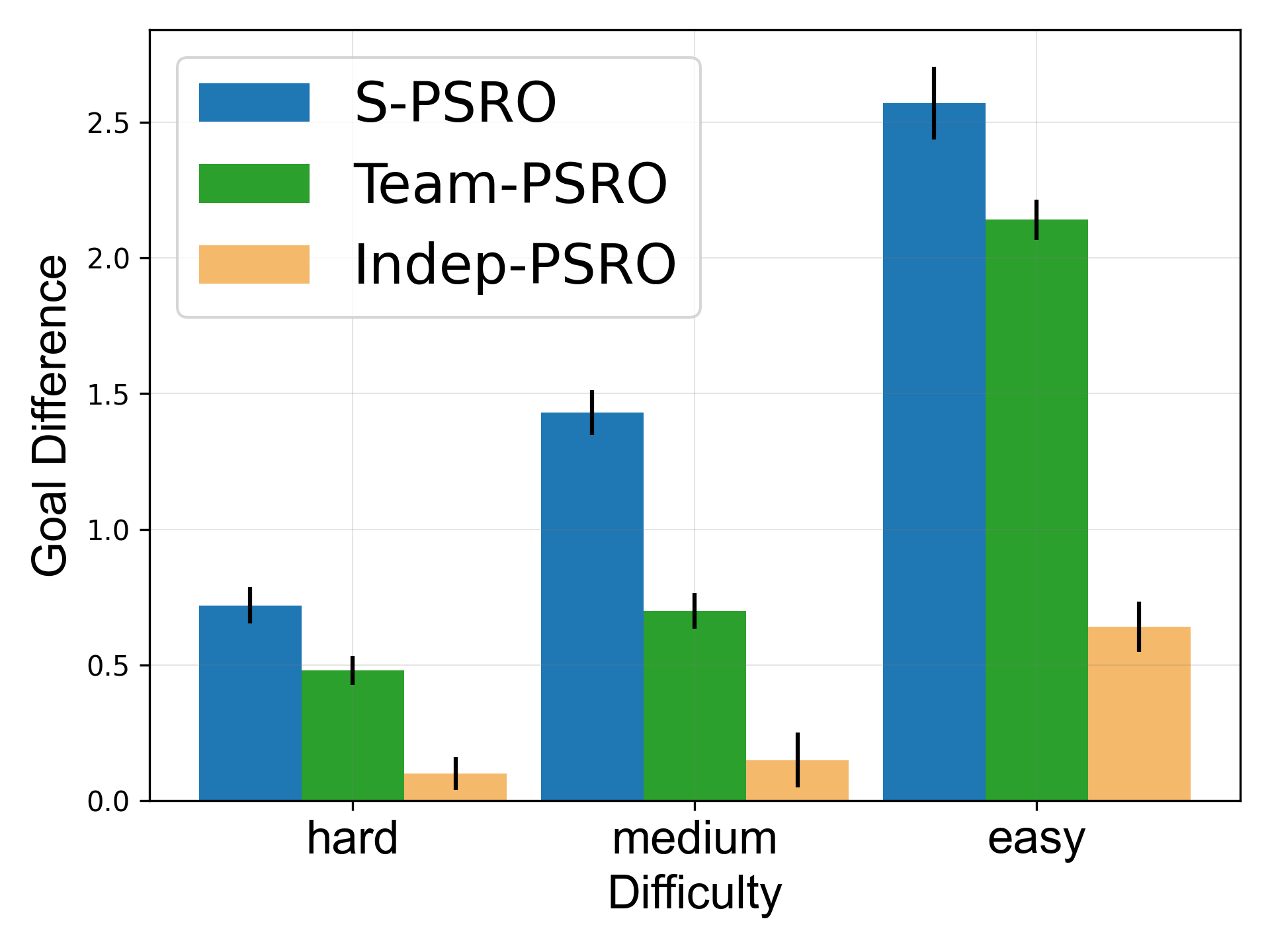}
        \label{fig:elo}
    }
    \subfigure[Performance Radar]{
        \includegraphics[width=0.44\columnwidth]{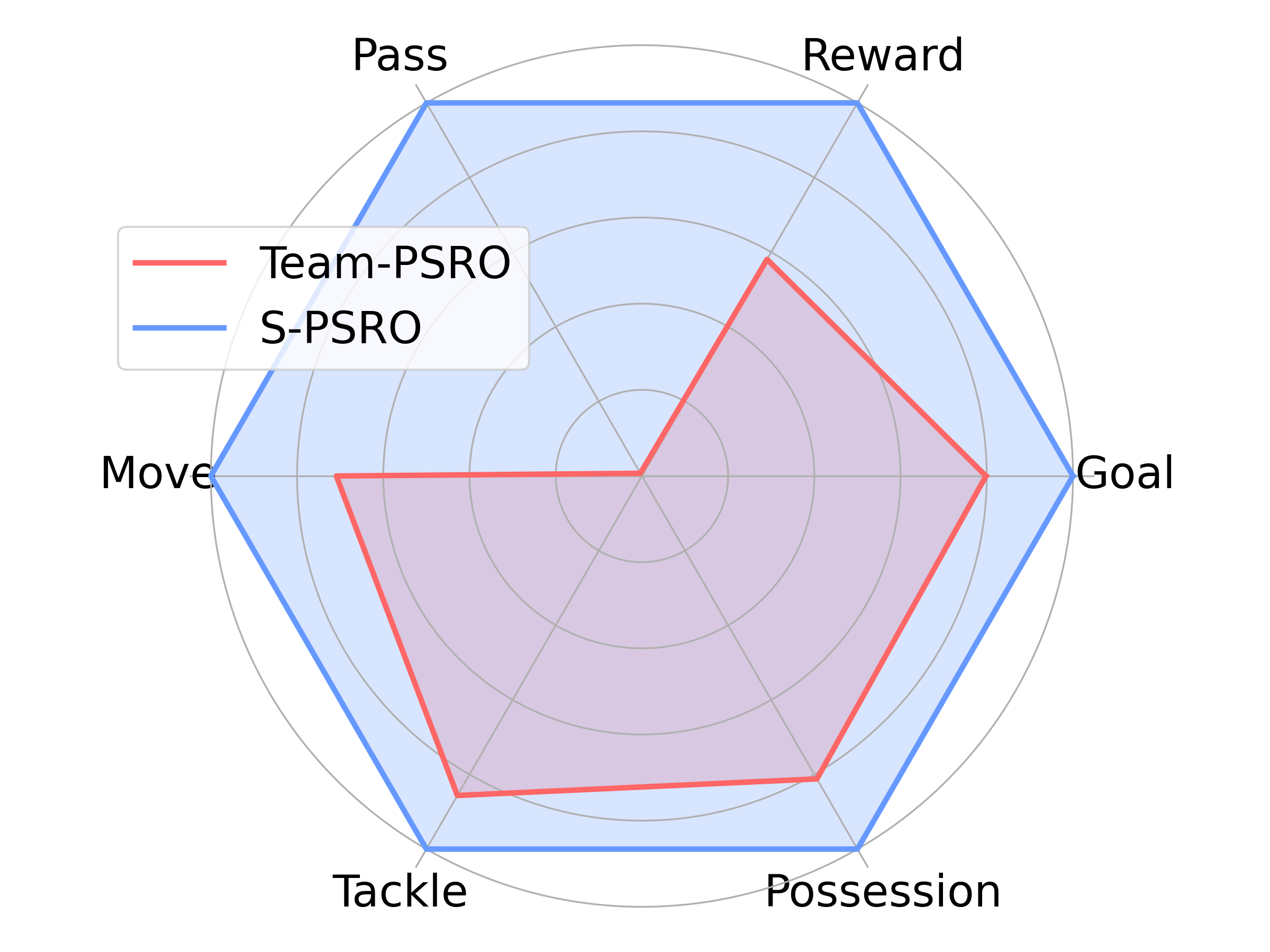}
        \label{fig:radar}
    }
    \subfigure[Relative Population Performance]{
        \includegraphics[width=0.58\columnwidth]{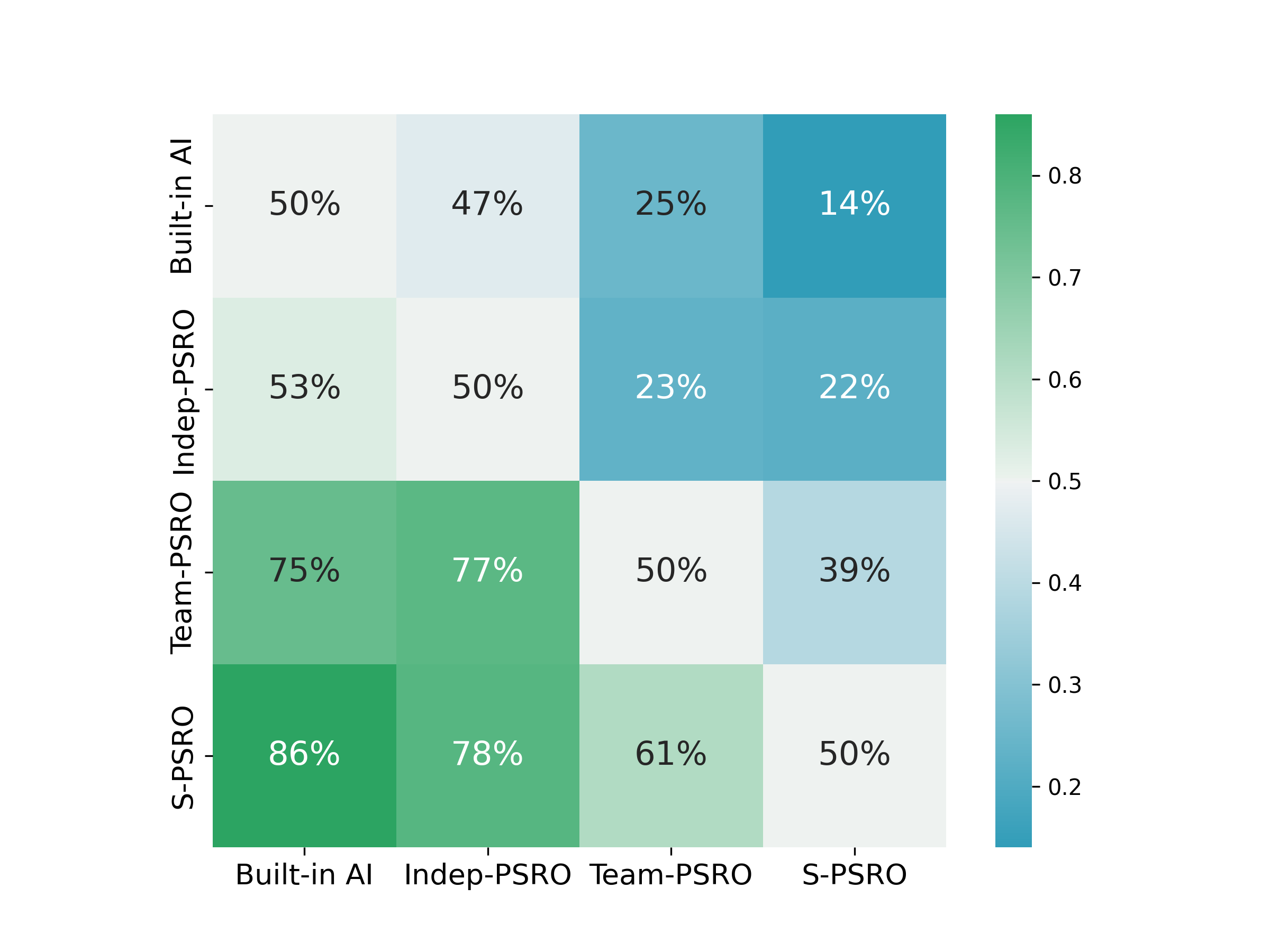}
        \label{fig:rpp}
    }
    
    \caption{Performance of S-PSRO, Team-PSRO and Indep-PSRO in Google Research Football, where S-PSRO surpasses the baselines.}
    \label{fig:grf}
\end{figure*}

\begin{table*}[t]
\caption{Performance of S-PSRO, Team-PSRO and PSRO in MAgent. MAgent is a gridworld battle scenario where each player has 21 actions. When increasing the number of teammates, the team joint action space explodes exponentially. We show in larger games (e.g., 12v12, 16v16), S-PSRO is capable of finding equilibrium policies with lower exploitability when confronting opponent teams with different exploitation ability.}
\label{tab:magent}
\vskip 0.15in
\begin{center}
\begin{small}
\begin{sc}
\scalebox{0.645}{
    \begin{tabular}{ccclllll}
        \toprule
        \multirow{2}*{Game Setting} & \multirow{2}*{Team Joint Action Space} & \multirow{2}*{Algorithm} &\multicolumn{5}{c}{Exploitability over Different Opponents}\\ 
        \cmidrule{4-8}
        && & Sequential Correlation & Joint Correlation &  Synchronized Correlation & No Correlation & Random\\
        \midrule 
        & & S-PSRO&9.520 (0.67)  &3.089 (0.23) & 6.575 (0.09) &3.405 (0.33) & -0.627(0)\\
        3v3 & 9.26e+\textbf{3} & Team-PSRO &7.251 (0.43)& 2.848 (0.28) & 5.743 (0.37)&3.117 (0.28) &-0.627 (0)\\
        & & \textbf{PSRO}&\textbf{2.428 (0.16)} &\textbf{2.122 (0.1)}&\textbf{0.625 (0.09)}&\textbf{1.888 (0.14)} & \textbf{-0.732 (0)}\\ 
        \midrule
        
        & & S-PSRO &20.223 (0.66) & 11.074 (0.48) &11.153 (0.56)&7.01 (0.43) &-4.640 (0)\\
        6v6 & 8.58e+\textbf{7} & Team-PSRO  &23.877 (0.73)&18.390 (0.62)&  13.581(0.56)&14.842 (0.61) & -2.980 (0)\\
        & & \textbf{PSRO} & \textbf{13.439 (0.56)}&\textbf{6.691 (0.33)}&\textbf{3.263 (0.11)}&\textbf{6.302 (0.26)} &\textbf{-5.377 (0)}\\ 

        \midrule
        
        & & \textbf{S-PSRO} & \textbf{12.964 (0.55)} &\textbf{-1.172 (0)}& \textbf{-2.062 (0.01)} & \textbf{0.403 (0.14)} & \textbf{-7.749 (0)}\\
        
        12v12 & 7.36e+\textbf{15} & Team-PSRO &28.182 (0.69) &4.931 (0.24)& 6.676 (0.32)& 16.060 (0.55) & -4.650 (0.01)\\
        
        & & PSRO &16.222 (0.55)& 2.488 (0.08)&2.138 (0.24)&7.418 (0.33)&  -4.992 (0) \\
        
        \midrule

         & &\textbf{S-PSRO} &\textbf{13.449 (0.43)} &\textbf{-1.711 (0.03)} & \textbf{-10.198 (0.09)} &  \textbf{-0.563 (0.24)}& \textbf{-6.854 (0.01)}\\
         
        16v16 & 1.43\textbf{e+21} & Team-PSRO &25.412 (0.60)& -1.454 (0.01) &-7.941 (0.13)& 16.767 (0.48)& -3.394 (0.01)\\
        
        & & PSRO & 26.929 (0.80)& 0.597 (0.04) &6.396 (0.37)& 22.239 (0.69)&-2.656 (0)\\
        
        \bottomrule
    \end{tabular}
    }
\end{sc}
\end{small}
\end{center}
\vskip -0.1in
\end{table*}

\section{Experiments}

In this section, we first valid the performance of S-PSRO with Team-PSRO \cite{TeamPSRO} and Online Double Oracle \cite{ODO} in a normal form game, where S-PSRO achieves lower exploitability. Then we compare rCTME under sequential correlation, with rCTME under pivot-followers correlation, CTME, NE in larger 2t0s games by approximate algorithms. Concretely, we use S-PSRO to find rCTME under sequential correlation, Team-PSRO \cite{TeamPSRO} to find rCTME under synchronized correlation, PSRO with joint team policy to find CTME, and Indep-PSRO (using IPPO to compute best response) to find NE \cite{IPPOConverge}. In large games, we first study a gridworld environment MAgent Battle \cite{magent}, where S-PSRO achieves the lowest exploitability over opponent team with different exploitation ability as game scaling. Then in the challenging two-team full game in Google Research Football \cite{football}, S-PSRO achieves higher relative performance over Team-PSRO, Indep-PSRO and hard Built-in AI opponent. S-PSRO achieves higher goal difference, Elo rating. Besides, S-PSRO achieves full capability in individual performance metrics as well as cooperative performance metrics, outperforming Team-PSRO significantly.

\subsection{Normal Form Game}
\begin{figure}[ht]
\vskip 0.1in
\begin{center}
\centerline{\includegraphics[width=0.65\columnwidth]{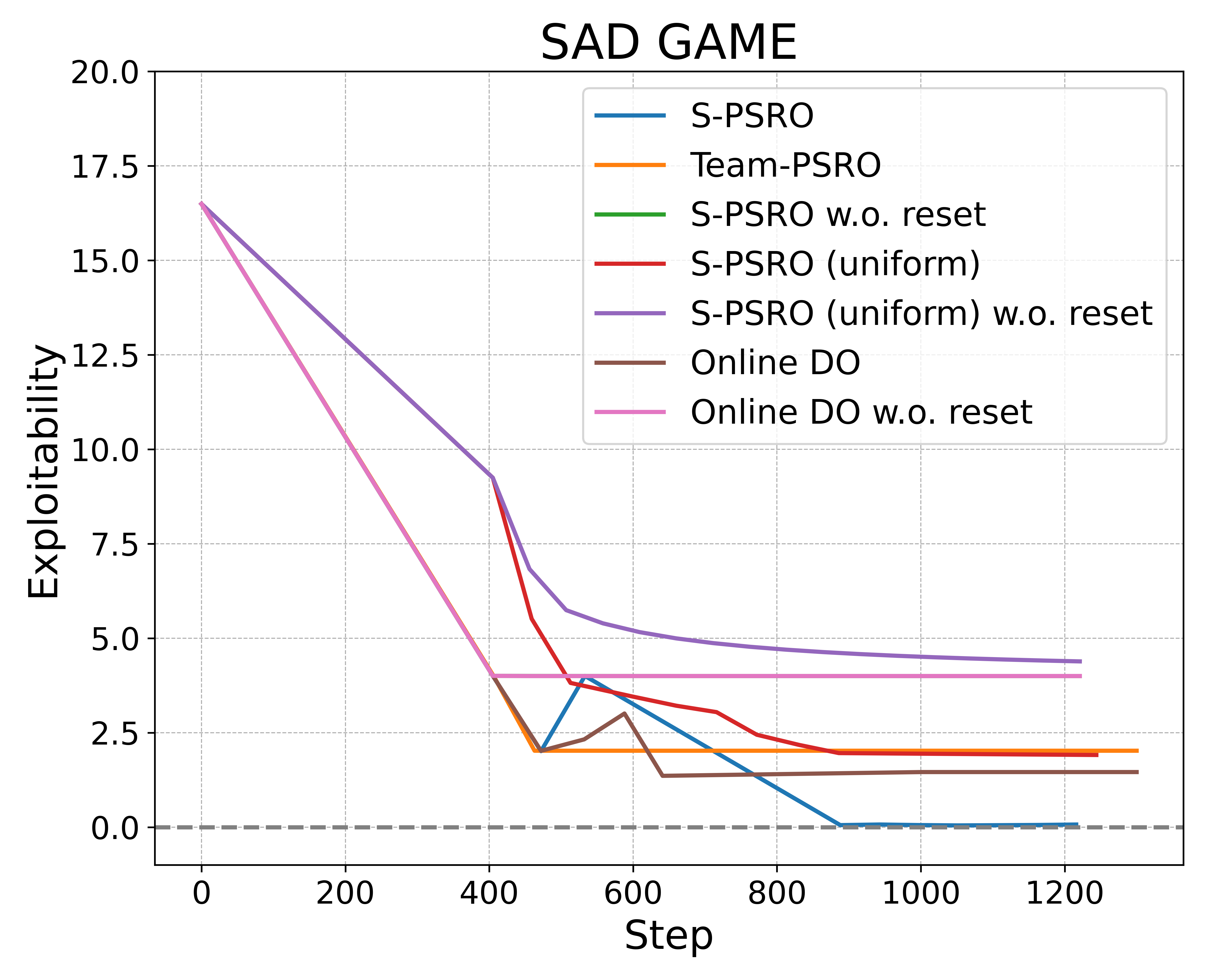}}
\caption{Exploitability in SAD games is calculated. }
\label{fig: sad}
\end{center}
\vskip -0.34in
\end{figure}

We validate the performance of S-PSRO in a matrix game with large action space, Seek-attack-defend (SAD) game \cite{FCP}. A seek-attack-defend (SAD) game consists of two teams of $N$ agents, each with the action space containing $A$ + 1 seeking action \{0, 1, 2, ..., $A$\} and two special actions \{\textit{attack}, \textit{defend}\}. We show the learning curve of exploitability of three learning based algorithms, including S-PSRO, Team-PSRO and Online Double Oracle in figure~\ref{fig: sad}, where S-PSRO eventually achieves lower exploitability than other algorithms.
\subsection{MAgent Battle}
MAgent Battle is a gridworld game where a red team of $N$ agents
fight against a blue team. At each step, agents can move to one
of the 12 nearest grids or attack one of the 8 surrounding grids
of themselves. To compare the scalability of S-PSRO, Team-PSRO and PSRO, we run algorithms in the 3-vs-3, 6-vs-6, 12-vs-12, 16-vs-16 MAgent Battle game. Since the exploitability cannot be exactly calculated in this games, we estimate the approximate exploitability of the final equilibrium policies by utilizing random policy and differently correlated Best Response as opponent team policies. The averaged results over 3 seeds are shown in Table~\ref{tab:magent}. We show that though in small scale 3-vs-3 and 6-vs-6 game, PSRO achieves lowest exploitability over S-PSRO and Team-PSRO, in larger scale 12-vs-12 and 16-vs-16 with significantly larger action (policy) space, S-PSRO achieves the lowest exploitability when confronting different correlated opponent policies over PSRO as well as Team-PSRO. It also shows that (1) joint policy fails to learn a more complex cooperative relationships among teammates, (2) the impact of restrictions on policy space and constraints of the cooperative method in Team-PSRO becomes more severe in large games.

\subsection{Google Research Football}

\label{sec:grf}

To further demonstrate the effectiveness of S-PSRO and its scalability to complex team games, we utilized Google Research Football (GRF) \cite{football} as our benchmark. Google Research Football environment is a simulation environment for real-world football games, where each game consists of 3000 steps. We conducted training and evaluation of our algorithm on the full 5 vs 5 game in GRF based on the benchmark \cite{grf_benchmark}, where the goalkeeper is controlled by built-in AI, while the rest of the four players are controlled by our model.

We valid the performance of S-PSRO by comparing the Elo Ratings \cite{elorating}, Goal Difference, Relative Population Performance metrics with Indep-PSRO and Team-PSRO. As the results in figure~\ref{fig:grf} show, S-PSRO achieves better performance than the baselines. Furthermore, we compare S-PSRO and Team-PSRO in terms of individual performance metrics (Goal Possession, Reward) and cooperative performance metrics (Pass, Move, Tackle) in Figure~\ref{fig:radar}, with S-PSRO showcasing superior performance over Team-PSRO in both individual and cooperative aspects, particularly in passing.

\section{Conclusion}
In this work, we focus on two-team zero-sum games. Our main contributions include the definition of a uniform equilibrium framework encompassing existing equilibria for two-team zero-sum games. Under the framework, we further define an efficient sequential correlation mechanism, and an approximate algorithm to approximate the unexploitable equilibrium in large games. Last, we validate the proposed algorithm in large games. 
\clearpage
\section{Impact Statements}
Our research significantly advances game theory by developing unexploitable equilibria in large team zero-sum games, offering a new framework for strategic interactions. We introduce the correlated-team maxmin equilibrium to overcome the challenges in large game settings and propose a novel restricted version for practical applicability. Our efficient sequential correlation mechanism and an algorithm for approximating these equilibria further enhance this approach. Empirically validated in large team games like Google Research Football, our technique outperforms existing methods and demonstrates lower exploitability in competitive scenarios.

\nocite{langley00}

\bibliography{icml2023}
\bibliographystyle{icml2023}

\newpage
\appendix
\onecolumn
\section{Algorithm}

\label{sec: s-psro}

\subsection{SeBR}

We implement the sequential correlation cooperative consensus by a Sequential Best Response Oracle (SeBR). However, decomposing the team optimization process into a sequential extensive-form learning process leads to temporal decoupling, which poses a specific challenge: how to make every agent's policy optimization contribute to the overall improvement of the team reward.

To tackle the above challenge, we introduce the following team advantage decomposition theorem \cite{happo, mat}, which ensures a consistent and monotonic enhancement while optimizing agents sequentially.

\begin{equation}
\label{eq: advantage}
\begin{aligned}
A^{i_{1:n}}_{\pi}(o, a^{i_{1:n}}) = \sum_{m=1}^{n}A_{\pi}^{i_m}(o, a^{i_{1: m-1}}, a^{i_m})
\end{aligned}
\end{equation}

Building upon the aforementioned theorem, SeBR computes a Best Response of distributed team policy given the fixed opposing team policy. Specifically, at each iteration players make decisions based on the current team policy ($\pi_1, \pi_2, \dots, \pi_n$) and transmit their decision preferences and rewards through communication channel $Q$. Then individual policies will be optimized sequentially with the decomposed advantage loss, which is computed according to equation~(\ref{eq: advantage}) with the information from $Q$.

The pesudocode for SeBR is shown in Algorithm~\ref{algo:sebr}.

\begin{algorithm}[H]
\caption{SeBR}
\label{algo:sebr}
\begin{algorithmic}[1]
\Require Meta Policy of Opponent Team $\Pi_O$, Communication Channel $Q$
\State Initialize distributed team policy $(\pi_1, \pi_2, \dots, \pi_n)$
\For{$i = 1, \dots, \text{MaxIter}$}
    \State empty Q
    \State TeamAction = Action($\pi_1, \pi_2, \dots, \pi_n, Q$)
    \State OppoAction = Action($\Pi_O$)
    \State $R_T$ = Reward$_T$(TeamAction, OppoAction, Q)
    \For{$m = 1, \dots, n$}
    \State Extract $A_1, ..., A_m$ and $R_T$ from $Q$
    \State $Loss_m$ = DecomposedAdvantage($R_T,$
    \State $\{A_1, ..., A_m\}$)
    \State $\pi_m$ = Optimize($\pi_m, Loss_m$)
    \EndFor
    \If{Converge}
    break
    \EndIf
\EndFor
\State \textbf{return} $\pi_1, \pi_2, \dots, \pi_n$
\end{algorithmic}
\end{algorithm}

\section{Cooperative Performance of Team-PSRO and S-PSRO in Google Research Football}
\begin{figure}[H]
\centering 

\begin{minipage}[b]{0.9\textwidth} 
\centering 
\includegraphics[width=0.32\textwidth]{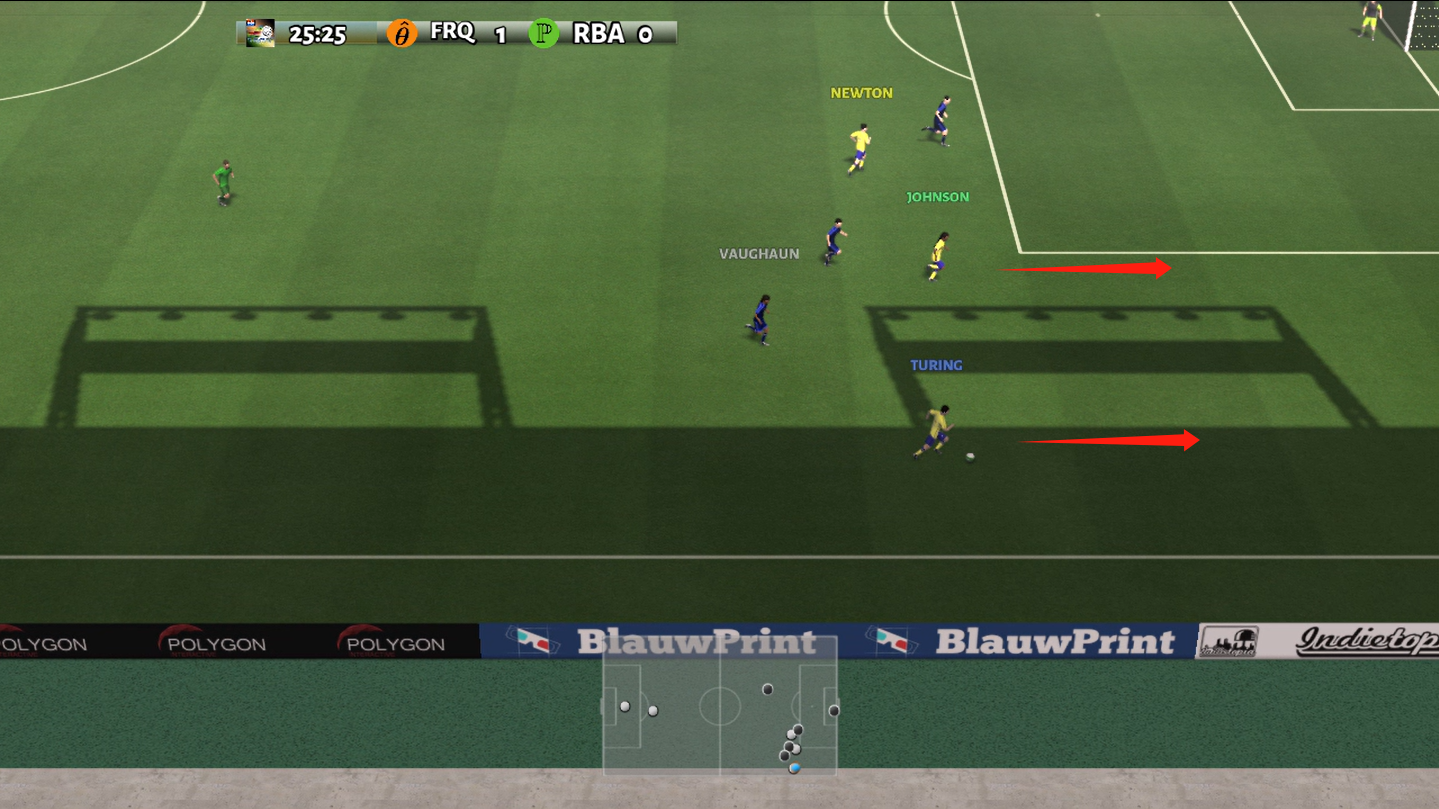}
\includegraphics[width=0.32\textwidth]{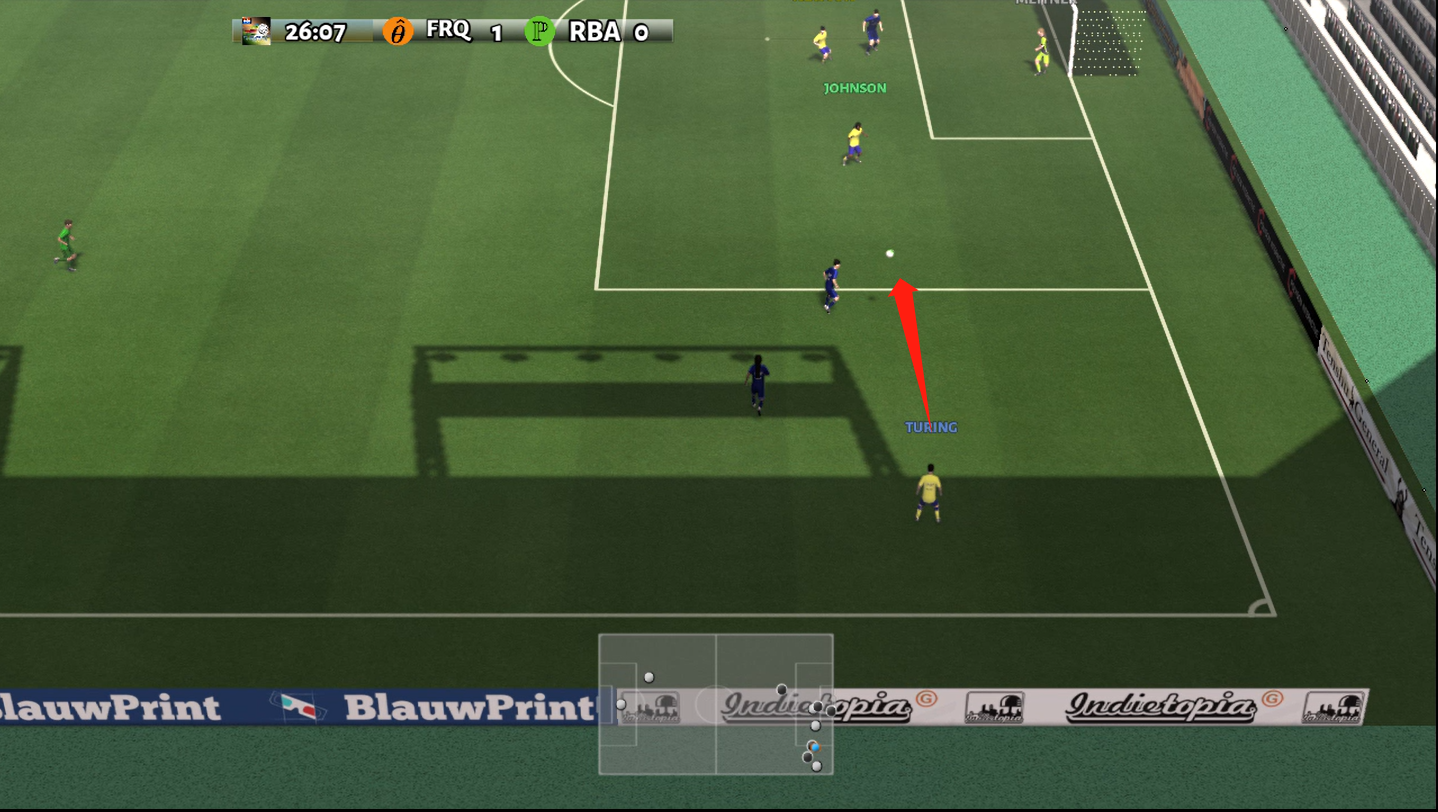}
\includegraphics[width=0.32\textwidth]{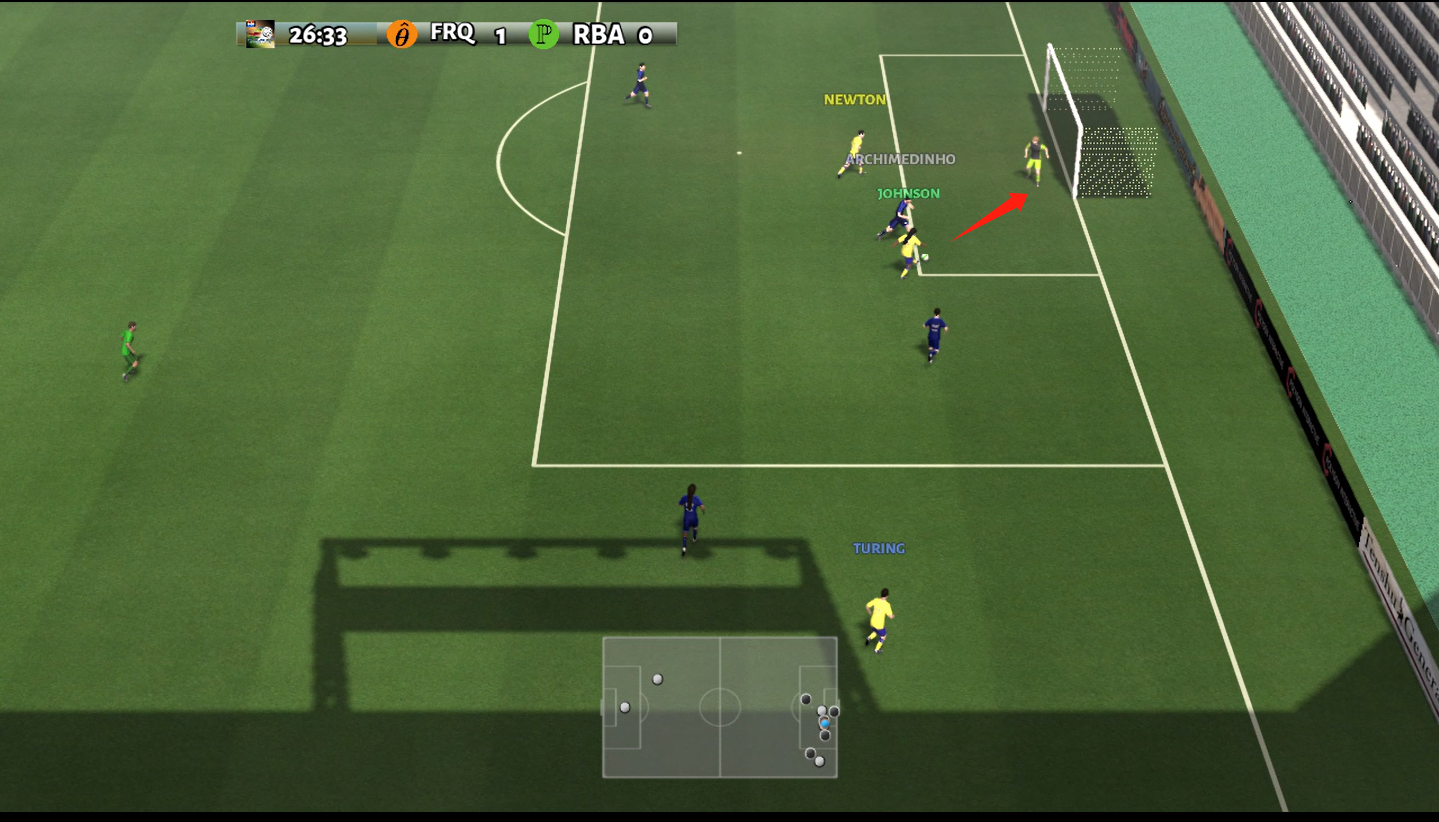}
\caption{Cooperative Behaviours of S-PSRO for Passing in Google Research Football}
\label{fig:s-psro passing}
\end{minipage}

\begin{minipage}[b]{0.9\textwidth} 
\centering 
\includegraphics[width=0.32\textwidth]{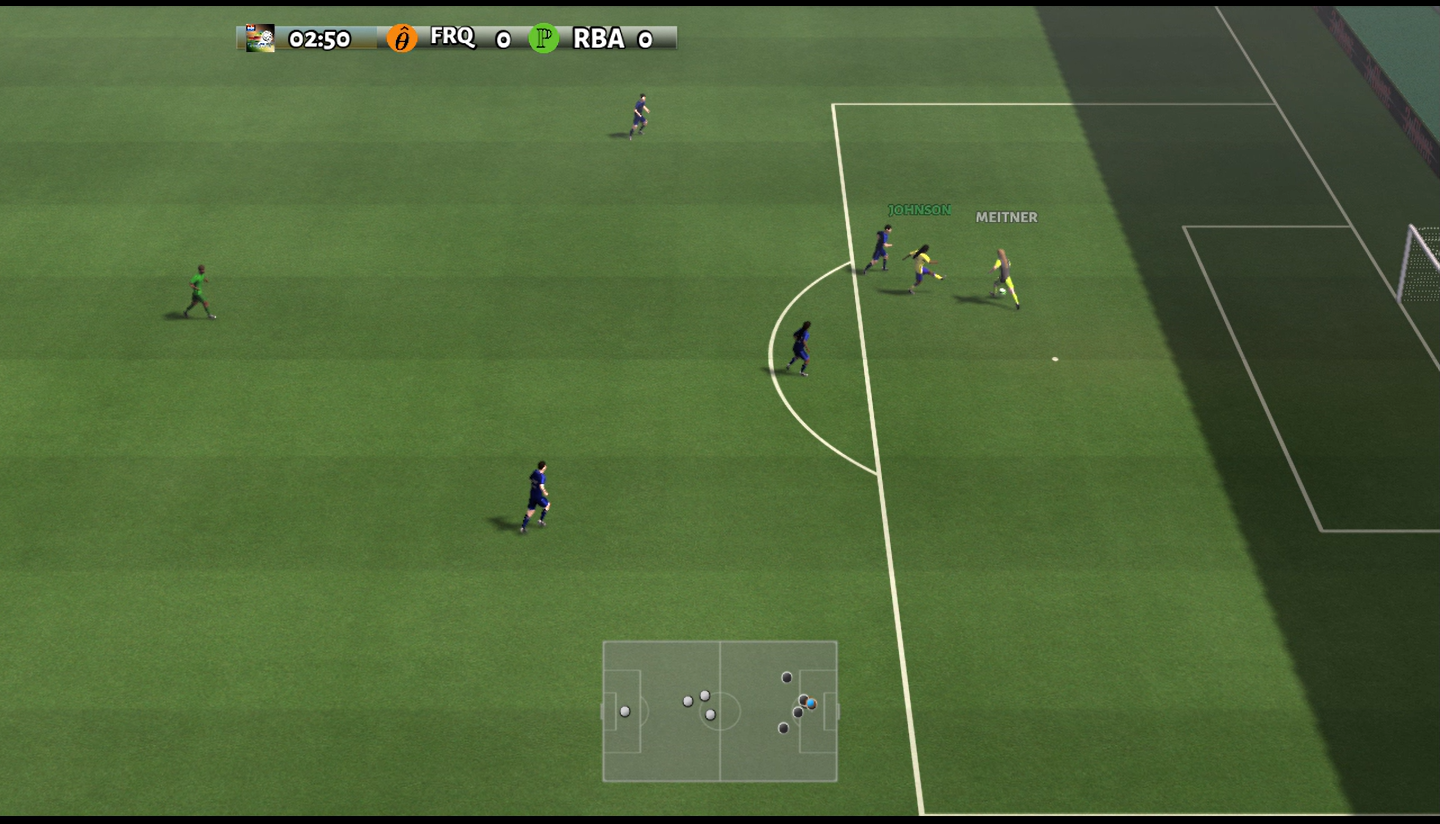}
\includegraphics[width=0.32\textwidth]{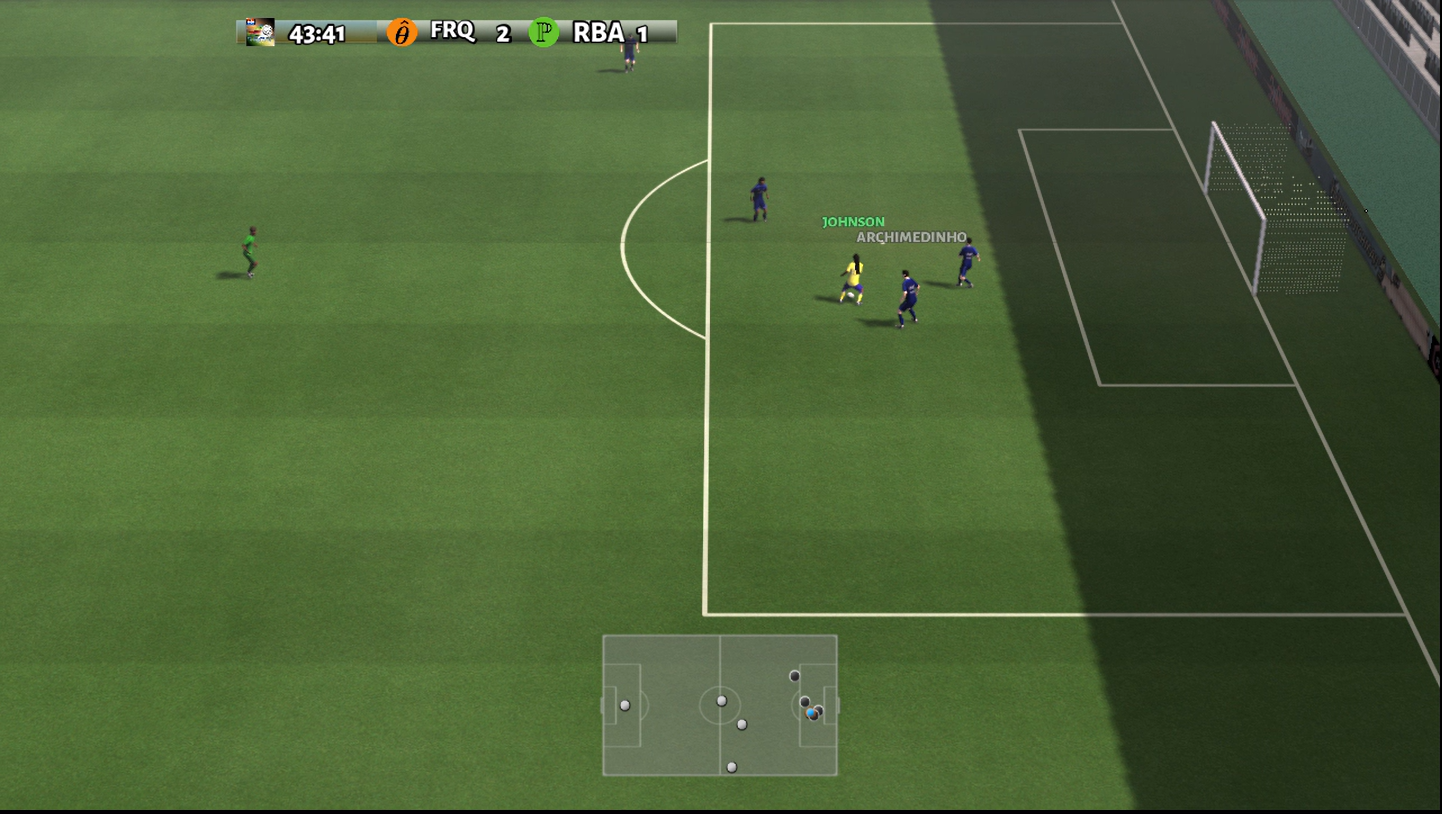}
\includegraphics[width=0.32\textwidth]{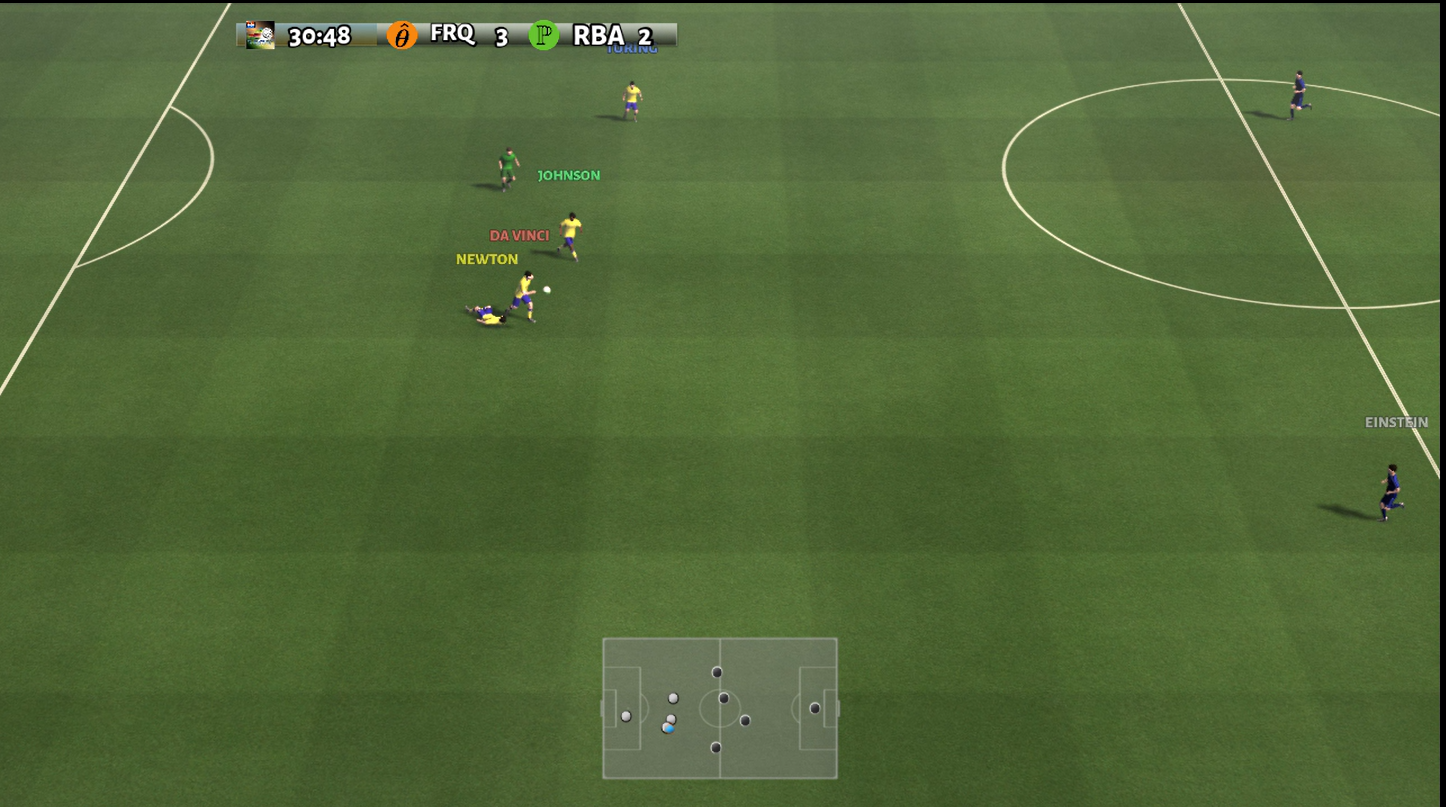}
\caption{Lack of Cooperation of Team-PSRO for Passing in Google Research Football}
\label{fig:team psro passing}
\end{minipage}
\end{figure}

We observe that, in comparison to pivot-followers consensus, sequential correlation consensus achieves a higher level of cooperation among teammates in complex tasks, such as passing in Google Research Football. 

As depicted in figure~\ref{fig:s-psro passing}, S-PSRO, an implementation of sequential correlation consensus in large-scale games, enables the player to successfully pass the football to their teammate. In contrast, in figure~\ref{fig:team psro passing}, Team-PSRO, an implementation of pivot-followers consensus in large-scale games, results in players' unsuccessful attempts to pass the football to their teammates.

\end{document}